# Quantification of nonlinear absorption in ternary As-Sb-Se chalcogenide glasses

P. Pradhan[1], Pritam Khan[2], J. R. Aswin[3], K. V. Adarsh[3], R. Naik[4], N. Das[1], A. K. Panda[1]

[1]Department of Electronics & Communication, NIST, Berhampur, 761008, Odisha

[2]Department of Physics, Kyushu University, Fukuoka 819-0395, Japan

[3]Department of Physics, IISER Bhopal, India

[4] Department of Physics, Utkal University, Bhubaneswar, 751004, India

*In this article, we studied intensity dependent third order nonlinear optical response in ternary $As_{40}Sb_7Se_{53}$ and $As_{40}Sb_{10}Se_{50}$ chalcogenide glasses by employing nanosecond Z-scan technique. At low intensity, we observed saturable absorption in $As_{40}Sb_7Se_{53}$ which makes a remarkable transition to reverse saturable absorption at higher intensities. On the other hand, when the Sb concentration increased in $As_{40}Sb_{10}Se_{50}$, saturable absorption disappears and the sample exhibits only two-photon absorption. Experimental results further indicate that the strong two-photon absorption in our samples can be exploited to fabricate high performance solid state optical limiting devices for next generation all-optical network.*

## I. Introduction

Understanding and characterizing the nonlinear optical properties play a significant role in device technologies because it provides added photonic functionality[1]. Therefore, quest for identifying the ideal optical material for fabricating nonlinear photonic devices remains a central topic among the researchers over the past few decades. Among the several materials, Chalcogenide glasses (ChGs), owing to their high third order nonlinear refractive index[2], exceptional infra-red (IR) optical transmittance[3], high photosensitivity [4, 5], high figure of merit (FOM) and low-cost easy synthesis [6] become a potential candidate to fulfil such requirements.

Among the family of ChGs, binary As-Se [7, 8] and ternary Ge-As-Se [6, 9, 10] system have been studied extensively to understand their third order nonlinear optical properties. The reason being the components Ge, As and Se have similar size and electronegativity, plus they have largest glass forming region and strong stability against crystallization [11-14]. Apart from the standard ChGs, recently new ChGs are synthesized to fabricate high-bit-rate optical processing i.e. switching system to operate at lower peak power. For example, ternary As–Sb–Se glasses are formed by substituting As atoms with Sb atoms in the As–Se system [15, 16]. It is believed that adding appropriate amount of Sb into the As-Se system could form good thermally stable ternary glasses with expanded glass forming region without altering the glass structure drastically [17]. Recent work on As-Sb-Se glasses has shown that with the increase in Sb content, specific heat of the system decreases which reduces the overall thermal entropy of the system to make the system more stable [15]. Although the linear optical properties of As-Sb-Se system has been studied extensively [15, 16], the nonlinear properties are hitherto unknown. Therefore, it is quite alluring to explore the third order nonlinear optical properties of As-Sb-Se glasses to exploit it for practical applications.

In this article, we employed nanosecond Z-scan technique to quantify the third order nonlinear optical properties of $As_{40}Sb_7Se_{53}$ and $As_{40}Sb_{10}Se_{50}$ glasses. Our results show a remarkable intensity mediated transition from saturable absorption (SA) to reverse saturable absorption (RSA) in $As_{40}Se_{53}Sb_7$ whereas with increase in SB content, SA disappears and $As_{40}Se_{50}Sb_{10}$ exhibits two-photon absorption (TPA).

## II. Experimental

Bulk $As_{40}Sb_7Se_{53}$ and $As_{40}Sb_{10}Se_{50}$ glasses were prepared by conventional melt quenching method in ice cooled water. Thin films of thickness 800 nm were prepared by conventional thermal evaporation in vacuum of about $5 \times 10^{-5}$ Torr at a fixed deposition rate of 2-5 A°/s. The Energy dispersive X-ray analysis indicates that the composition of the thin films matches well with that of the bulk ChGs.

To quantify the nonlinear optical response of the samples we employed open aperture Z-scan technique. In our measurement, 1064 nm, 5 ns pluses of Nd:YAG laser was used to excite the sample. Precautions were taken to avoid the sample heating and damage by using a fixed repetition rate of 10 Hz.[6] The Rayleigh length and the beam waist in our experiment were 1.6 mm and 17 μm, respectively. A 20 cm plano-convex lens is used to focus the Gaussian beam at the zero position of z axis of computer-controlled translation stage.

## III. Results and Discussions

To elucidate the optical properties, first we recorded the optical transmission spectra of the samples as shown in Fig. 1(a). Clearly, in the high absorbing region, optical absorption edge blue shifted when Sb concentration increases from 7 to 10 % which results in the increase in bandgap ($E_g$). We used classical Tauc plot to quantitatively determine $E_g$ of the samples following Tauc's relation [18]

$$\alpha h\vartheta = B(h\vartheta - E_g)^2 \qquad (1)$$

where hv is the photon energy, α is the absorption coefficient, and B is band tailing parameter. The intercept of the straight line at the photon energy axis will provide $E_g$ as shown in Fig. 1(b). From the best fit of the experimental data by using Eq. (1), we found that $E_g$ is 1.728 ± 0.003 eV and 1.752 ± 0.002 eV for $As_{40}Sb_7Se_{53}$ and $As_{40}Sb_{10}Se_{50}$ respectively.

After understanding the linear optical properties and calculating the bandgap of the samples, we aim to quantify the third order optical nonlinearity by employing open aperture Z-scan technique. In this regard, Fig. 2(a) and (b) represents series of open aperture traces of normalized Z-scan peak shapes at different peak intensities of off-resonant excitations for $As_{40} Sb_7Se_{53}$ and $As_{40} Sb_{10}Se_{50}$ respectively. It is important to note that for both the samples, the peak intensities are measured at the focal point, i.e. z = 0. Let us first discuss $As_{40} Sb_7Se_{53}$ sample. It can be seen from Fig. 2(a) that at a reasonably lower intensity of 11MW/cm$^2$, normalized transmittance of the sample increases gradually with the sample position approaching focal point while maximum transmittance is observed at z = 0. Clearly, this response is typical for saturable absorption (SA) observed for $As_{40} Sb_7Se_{53}$ at lower intensity. In a stark contrast, when the intensity increases from 11 to 13, 32 and 37 MW/cm$^2$, the normalized transmittance exhibits strong nonlinear absorption near the focal point, i.e. transmission increases gradually at the far field |z| >2 cm. Clearly, such response replicates the observation of reverse saturable absorption (RSA) in $As_{40}Sb_7Se_{53}$ at moderately higher intensities. Therefore, our result demonstrates a remarkable transition from SA to RSA in $As_{40}Sb_7Se_{53}$ with the increase in input intensity.

On the other hand, when the Sb concentration increases, i.e. in $As_{40}Sb_{10}Se_{50}$, we did not observe any non-linear optical response below the threshold peak intensity of 25 MW/cm$^2$. For example, at 11 MW/cm$^2$ normalized transmittance exhibits a flat line in Fig. 2(b), indicating that no effect takes place at such lower intensity. At 25 MW/cm$^2$ the normalized transmittance exhibits nonlinear absorption, manifested by a dip at the focal point. As the band gap of $As_{40}Sb_{10}Se_{50}$ is above the single photon absorption energy of 1064 nm, it is reasonable to assume that the non-linear absorption in $As_{40}Sb_{10}Se_{50}$ is associated with two-photon absorption (TPA). Interestingly, the decrease in transmission, i.e. non-linear absorption increases with the increase in input intensity from 25 to 47 MW/cm$^2$ which further validates the observation of TPA in $As_{40}Sb_{10}Se_{50}$ [19].

The observed effects of SA, RSA and TPA in our samples can be explained by a three-level model, characterized by valence band, conduction band and the defect states which is intrinsic to ChGs [2]. As already mentioned, 1064 nm excitation lies close to the defect states which is well below the bandgap of the samples. Naturally, such excitation creates resonant single photon absorption to the defect state which will eventually give rise to SA. At lower intensities, the reason for observing SA is two-fold: first, density of the defect states is low which are almost occupied and second, Pauli blocking hinders any further absorption. Consequently, we observe SA in $As_{40}Sb_7Se_{53}$ at 11 MW/cm$^2$. The RSA at higher intensities can be explained on the basis of TPA arising from band to band transition. In a two-level system, electron simultaneously absorbs two photons of 1064 nm and makes a transition from valence to conduction band which gives rise to RSA in our sample.

After explaining the observed non-linear effects, it is of tremendous importance to quantify such effects to exploit for practical applications. In this regard, we fitted the experimental data using the differential equation by taking care both SA and TPA [20].

$$\frac{dI}{dz} = -\alpha(I)I \qquad (2)$$

Here (I) is and z are the intensity and propagation distance inside the sample; whereas α(I) can be written as:

$$\alpha(I) = \frac{\alpha_0}{1+\frac{I}{I_S}} + \beta_{TPA}I \qquad (3)$$

Where $\alpha_0$, β and $I_s$ are the linear absorption coefficient, TPA coefficient, and saturation intensity respectively. Eq. [2] can be solved for Gaussian ultrafast beam and the normalized transmittance as a function of position z in our Z-scan measurement can be expressed as:

$$T_N = \frac{1}{q_0\sqrt{\pi}} \int_{-\infty}^{+\infty} \ln(1 + q_0 e^{-t^2})\, dt \qquad (4)$$

Where $q_0 = \frac{\beta I_0 L_{eff}}{1+\frac{z^2}{z_0^2}}$ and $L_{eff} = \frac{(1-e^{-\alpha L})}{\alpha}$ Were $I_0$, $z_0$, L, and $L_{eff}$ are the peak intensity at the focus (z = 0), Rayleigh length, sample thickness, and effective length respectively. We calculated. β from the best fit to the normalized transmittance and are shown in Fig. 2(c). It is quite evident from the figure that for both the samples β, i.e. TPA coefficient increases, however with different rate. For example, in $As_{40}Sb_7Se_{53}$, β increases from (2.75 ± 0.38) ×$10^5$ to (6.98 ± 0.27) × $10^5$ cm/GW when input intensity raised from 11 to 37 MW/cm$^2$. Such strong increase in β reflects RSA process. On the other hand, β changes from (6.46 ± 0.35) ×$10^5$ to (7.41 ± 0.25) × $10^5$ cm/GW when input intensity changes from 25 to 47 MW/cm$^2$ and such behavior is signature for TPA. We presume that the enhancement of β as a function of input intensity in both the samples make them an ideal candidate for designing optical limiting devices. At this point, we have compared the TPA (β) obtained in our study with that of the results obtained before for other ChGs and are shown in Table I.

| Sample | $\lambda_{ex}$(nm) | $\beta$ (cm/GW) | References |
| --- | --- | --- | --- |
| $As_{40}Sb_7Se_{53}$ | 1064 | $(6.98 \pm 0.27) \times 10^5$ | Present work |
| $As_{40}Se_{50}Sb_{10}$ | 1064 | $(7.41 \pm 0.25) \times 10^5$ | Present work |
| $As_{40}Se_{60}$ | 1150 | 0.01 | [9] |
| $As_{40}S_{60}$ | 1250 | 2.8 | [21] |
| $As_{40}S_{40}Sb_{20}$ | 1250 | 0.22 | [21] |
| $Ge_{15}Sb_{10}Se_{75}$ | 1150 | 1.27 | [9] |

**Table I**. Comparison of TPA coefficients of As-Sb-Se chalcogenide glasses with some other ChGs systems.

The fruitfulness of an experimental results lies within the application to practical purposes. Therefore, it is important to use the present results to design a device, which is optical limiter in our case. It is well known that an optical limiting device allows the low intensity light beam while blocking the high intensity beams. Optical limiter plays an important role in protecting optoelectronic detectors as well as human eye from intense laser sources. A schematic diagram of a solid-state optical limiter is shown in Fig. 3 based on thin films of $As_{40}Sb_{10}Se_{50}$ on glass substrate. The working principle of an optical limiter is based on the fact that transmission decreases when the input laser intensity is above the certain threshold. Therefore, our samples are ideal candidate because it is quite evident from Fig. 4(a) and (c) that normalized transmittance decreases dramatically as a function of input intensity which is the primary criterion to fabricate an optical limiter. There are two important parameters to evaluate the performance of an optical limiter, namely onset threshold intensity $F_{ON}$ and optical limiting threshold intensity $F_{OL}$. $F_{ON}$ and $F_{OL}$ are defined as the input intensity at which the normalized transmittance deviated from linearity and drops below 50 % respectively. At this moment, by comparing Fig. 4(a) and (c) we learnt although for the both the samples transmission decreases, only for the sample with higher Sb content, i.e. $As_{40}Sb_{10}Se_{50}$ it reaches below 50 % to fulfil the criteria for optical limiter. At this moment, it is important to note that an ideal optical limiter should have very low value of both $F_{ON}$ and $F_{OL}$ [22]. Fig. 4(c) provides a useful tool to calculate both $F_{ON}$ and $F_{OL}$. Consequently, we found that for $As_{40}Sb_{10}Se_{50}$, $F_{ON}$ and $F_{OL}$ are 4.7 MW/cm$^2$ (0.0379 J/cm$^2$) and 38 MW/cm$^2$ (0.266 J/cm$^2$) respectively. To demonstrate the suitability of our samples as an optical limiter, we have plotted in Fig. 4(b) and (d) the output fluence as a function of input fluence. Quite clearly, for both the samples at lower intensity, $F_{out}$ is proportional to $F_{in}$, following Beer-Lambart law and consequently our device remains inactive in linear region as shown in Fig. 3. However, at higher input fluence, we observed that the limiting differential transmittance ($F_{out}/F_{in}$) shows a deviation at from the straight line i.e. the output intensity is limited by the material for all higher input intensities, i.e. our device becomes active optical limiter.

## IV. Conclusions

In conclusion, we have quantified third order nonlinear optical response in As-Sb-Se ChGs by employing nanosecond Z-scan technique. Intensity dependent Z-scan measurements demonstrates a remarkable transition from SA to RSA in $As_{40}Sb_7Se_{53}$, in a stark contrast to the observed TPA in $As_{40}Sb_{10}Se_{50}$. The strong enhancement of β from $(2.75 \pm 0.38) \times 10^5$ to $(6.98 \pm 0.27) \times 10^5$ cm/GW as a function of intensity validates the observation of RSA in $As_{40}Sb_7Se_{53}$. The observed nonlinear effects are explained by assuming a three-level model characterized by valance band, conduction band and defect states of ChGs. Apart from that, we also showed that the strong TPA in $As_{40}Sb_{10}Se_{50}$ can be effectively used to fabricate solid state optical limiting device for all-optical network with efficient device parameters.

## Acknowledgements

The authors thank Department of Science and Technology (DST), Govt of India for DST-INSPIRE Research grant of Dr. Naik and Physics Department, IISER, Bhopal for Z scan and optical measurements.

# References

[1] T. Kuriakose, E. Baudet, T. Halenkovič, M. M. R. Elsawy, P. Němec, V. Nazabal, G. Renversez, and M. Chauvet, Opt. Commun. **403**, 352 (2017).

[2] D. Mandal, R. K. Yadav, A. Mondal, S. K. Bera, J. R. Aswin, P. Nemec, T. Halenkovic, and K. V. Adarsh, Opt. Lett. **43**, 4787 (2018).

[3] S. Dai, F. Chen, Y. Xu, Z. Xu, X. Shen, T. Xu, R. Wang, and W. Ji, Opt. Express **23**, 1300 (2015).

[4] P. Khan, A. R. Barik, E. M. Vinod, K. S. Sangunni, H. Jain, and K. V. Adarsh, Opt. Express **20**, 12416 (2012).

[5] P. Khan, T. Saxena, H. Jain, and K. V. Adarsh, Sci. Rep. **4**, 6573 (2014).

[6] A. R. Barik, K. V. Adarsh, R. Naik, C. S. S. Sandeep, R. Philip, D. Zhao, and H. Jain, Appl. Phy. Lett. **98**, 201111 (2011).

[7] K. Ogusu and K. Shinkawa, Opt. Express **17**, 8165 (2009).

[8] S. Y. Kim, M. Kang, and S. Y. Choi, Thin Solid Films, **493 (1-2),** 207 (2005).

[9] T. Wang, X. Gai, W. Wei, R. Wang, Z. Yang, X. Shen, S. Madden, and B. Luther-Davies, Opt. Mat. Express **4**, 1011 (2014).

[10] J. T. Gopinath, M. Soljačić, E. P. Ippen, V. N. Fuflyigin, W. A. King, and M. Shurgalin, J. Appl. Phys. **96**, 6931 (2004).

[11] P. Khan, T. Saxena, and K. V. Adarsh, Opt. Lett. **40**, 768 (2015).

[12] L. Calvez, Z. Yang, and P. Lucas, Phys. Rev. Lett. **101**, (2008).

[13] P. Khan, A. Bhattacharya, A. Joshy, V. Sathe, U. Deshpande, and K. V. Adarsh, Thin Solid Films **621**, 76 (2017).

[14] P. Lucas, J. Phys.: Condens. Matter **18**, 5629 (2006).

[15] R. Naik, A. Jain, E. M. Vinod, R. Ganesan, and K. S. Sangunni, Phys. Status Solidi C **8**, 2785 (2011).

[16] R. Naik, A. Jain, R. Ganesan, and K.S. Sangunni, Thin Solid Films **520**, 2510 (2012).

[17] D. Garcı́a-G. Barreda, J. Vázquez *, P.L. López-Alemany, P. Villares, and R. Jiménez-Garay, J. Phys. Chem. Solids **66**, 1783 (2005).

[18] J. Tauc, Amorphous and Liquid Semiconductors, Plenum Press, NewYork, USA,1979.

[19] A. K. Rana, Aneesh. J, Y. Kumar, A. M. S, K.V. Adarsh, S. Sen, and P.M. Shirage, Appl. Phys. Lett. **107**, 231907 (2015).

[20] M. Sheik-Bahae, A. A. Said, T. -H. Wei, D. J. Hagan, and E. W. Van Stryland, IEEE J. Quantum Electron. **26**, 760 (1990).

[21] J. M. Harbold, F. Ö. Ilday, F. W. Wise, J. S. Sanghera, V. Q. Nguyen, L. B. Shaw, and I. D. Aggarwal, Opt. Lett. **27**, 119 (2002).

[22] R.K. Yadav, J. Aneesh, R. Sharma, P. Abhiramnath, T.K. Maji, G.J. Omar, A.K. Mishra, D. Karmakar, and K.V. Adarsh, Phys. Rev. Applied **9**, (2018).

**Figure captions**

**Fig. 1.** (a) Optical transmission spectra of $As_{40}Sb_7Se_{53}$ and $As_{40}Sb_{10}Se_{50}$ thin films and (b) Tauc plot used for calculating the bandgap of the respective samples.

**Fig. 2.** Normalized transmittance as a function of position in open aperture Z-scan for 5 ns, 1064 nm excitation at different peak intensities for (a) $As_{40}Sb_7Se_{53}$ and (b) $As_{40}Sb_{10}Se_{50}$ thin films (c) Quantification of TPA coefficient β as a function of peak intensity for the two samples.

**Fig. 3.** Schematic representation of a solid-state based passive optical limiting device consisting of a thin film with a linear transmittance of 50 %. At the violet region, our device allows to transmit the low intensity light. However, in the yellow region when the high intensity beam falls, our device acts as an optical limiter and attenuates the intense beam. Here $I_1$, $I_2$ and $I_3$ are below the threshold intensity and $I_4$, $I_5$ $I_6$, and $I_7$ are above it.

**Fig. 4.** Normalized transmittance as a function of input intensity for (a) $As_{40}Sb_7Se_{53}$ and (c) $As_{40}Sb_{10}Se_{50}$ respectively. Here the solid symbols and lines represent the experimental data and theoretical fit respectively. Optical limiting curve represented by the variation of output intensity as a function of input intensity for (b) $As_{40}Sb_7Se_{53}$ and (d) $As_{40}Sb_{10}Se_{50}$ respectively. The black line shows the linear transmittance and the curved lines represent the theoretical fit.

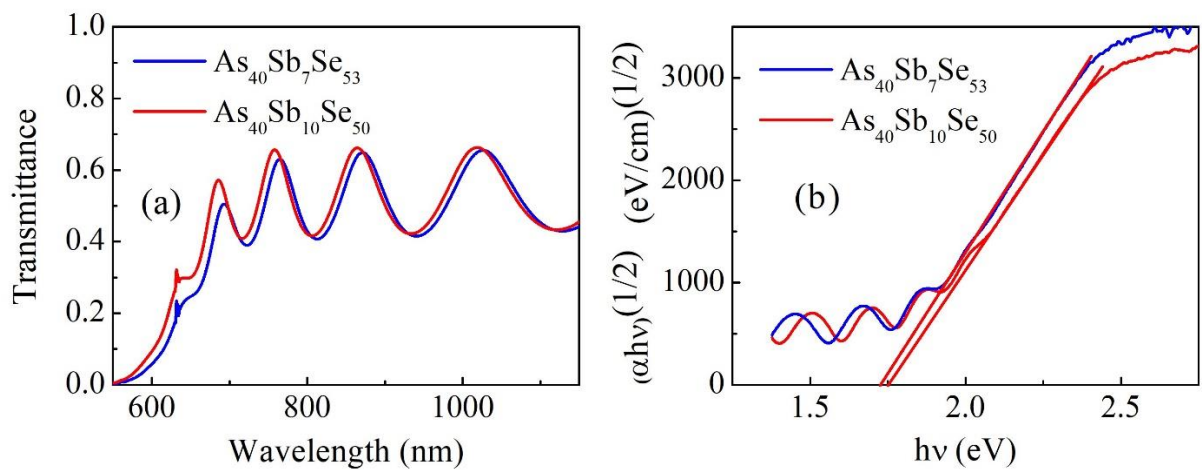

Fig. 2

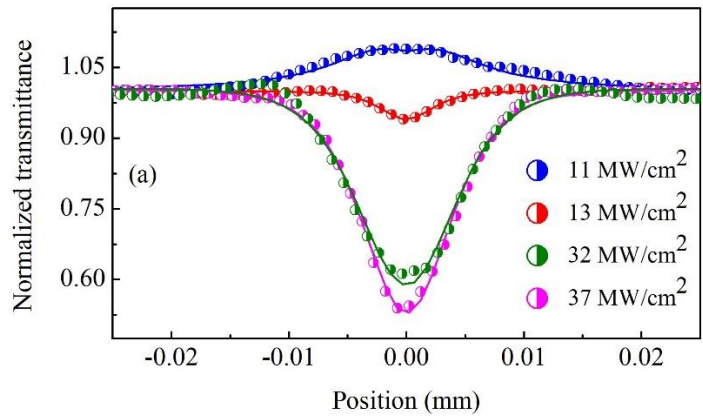

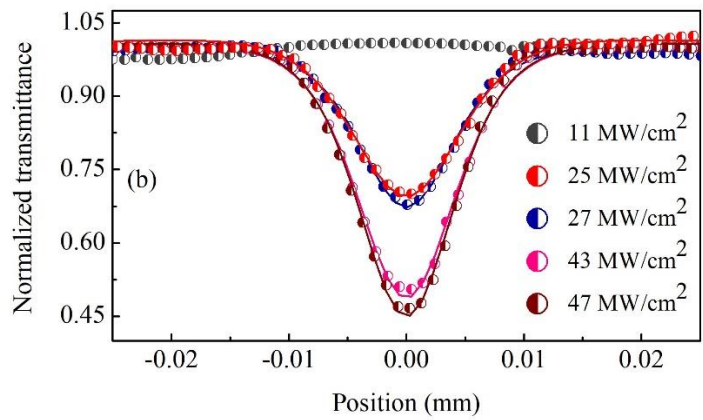

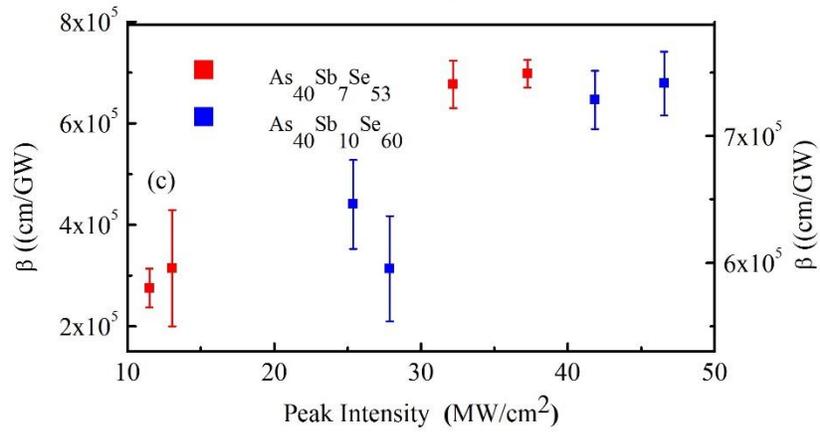

**Fig. 3**

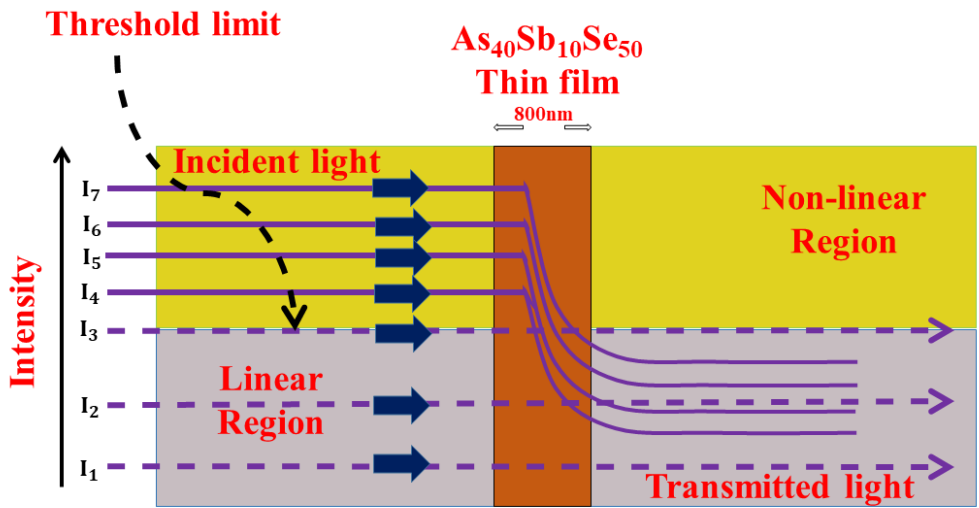

**Fig. 4**

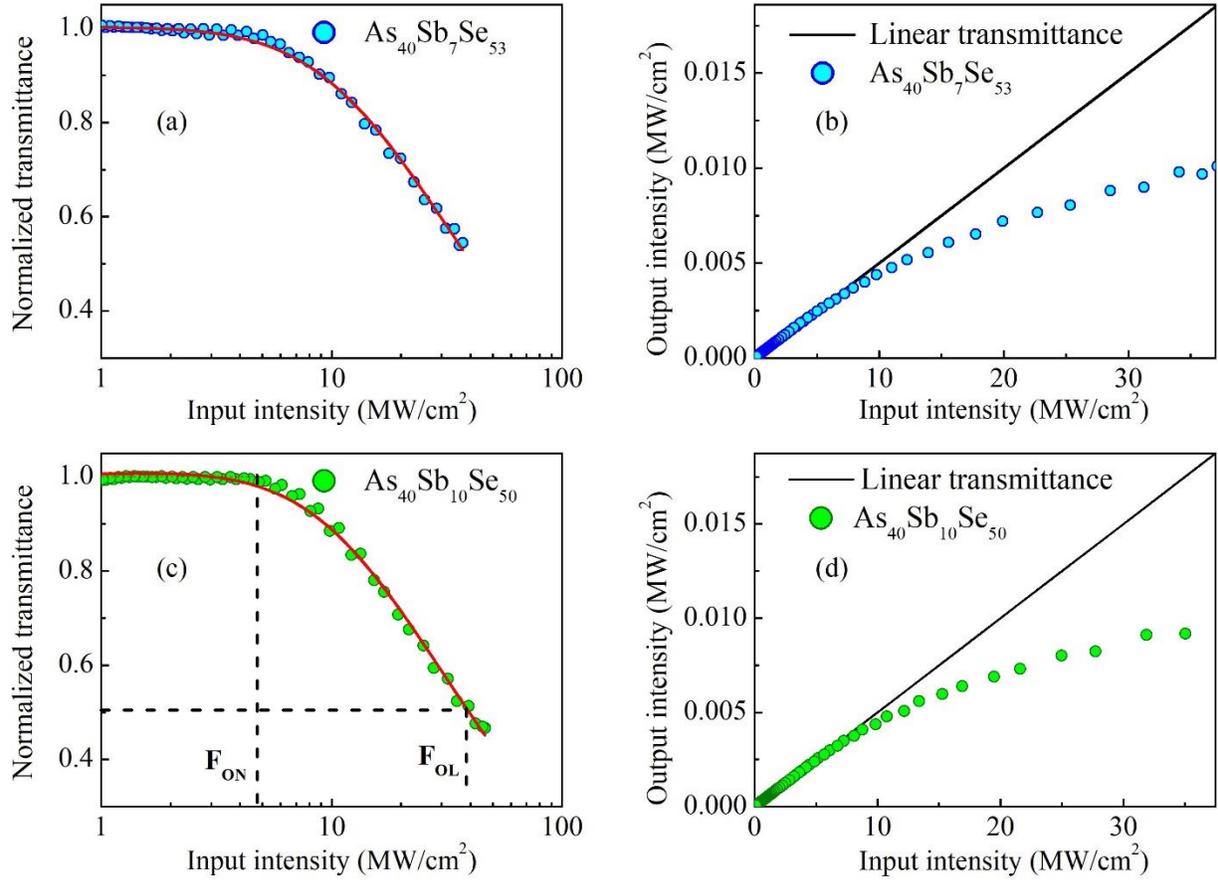